\documentclass[preprints,article,accept,moreauthors,pdftex]{Definitions/mdpi}
\firstpage{1} 
\pubvolume{6}
\issuenum{4}
\articlenumber{70}
\pubyear{2022}
\copyrightyear{2022}
\externaleditor{Academic Editors: Fabrizio Salvatore,
Alessandro Cerri, Antonella De Santo
and Iacopo Vivarelli} %MDPI: plaese add Academic Editor
\datereceived{21 August 2022} 
\dateaccepted{18 October 2022} 
\datepublished{27 October 2022} 
\hreflink{https://doi.org/\\ 10.3390/instruments6040070} % If needed use \linebreak
\pdfoutput=1
\usepackage{siunitx}
\usepackage{float}
\usepackage{changes}

\Title{Design and Test-Beam Results of the FoCal-H Demonstrator~Prototype}

\TitleCitation{Design and Test-Beam Results of the FoCal-H Demonstrator Prototype}

\Author{{Radoslav Simeonov} \orcidA{} on behalf of the ALICE Collaboration} 
\AuthorNames{Radoslav Simeonov}

\AuthorCitation{Simeonov, R., on behalf of the ALICE Collaboration.}
\address[1]{Faculty of Physics, Sofia University ``St. Kliment Ohridski'', 5 J. Bourchier Blvd, 1164 Sofia, Bulgaria;  radoslav.simeonov@cern.ch \\
}

\abstract{The forward calorimeter (FoCal) of ALICE, planned to be operational for LHC Run 4, will cover the pseudorapidity range 3.4  $\leq \eta \leq $  5.8 allowing to probe the unexplored region of Bjorken-x down to $10^{-6}$. The hadronic section of the FoCal (FoCal-H) will be based on copper capillary tubes and scintillating fibers inside, with light read out by silicon photomultipliers (SiPM). A “proof of concept” demonstration prototype was built and tested in the H6 beamline at the CERN SPS in the beginning of October, 2021, exposing it to an unseparated charged particle beam with energy in the interval 20 GeV--80 GeV. The design of the prototype as well as the results of the energy reconstruction are presented and the validation with a GEANT4-based simulation is 
discussed. }


\begin{document}

\section{Introduction}
The ALICE experiment at the CERN LHC is devoted to the study of the physics properties of strongly interacting matter and quark-gluon plasma at extreme values of energy density and temperature in nucleus-nucleus collisions \cite{ALICE:2008ngc}. 
An ALICE upgrade phase, to be completed before LHC Run 4, foresees an introduction of a new forward calorimeter detector (FoCal) \cite{Colella:2022lmb}.

The FoCal is intended to cover the pseudorapidity 
region $3.4 \leq \eta \leq 5.8$, which will enable measurements 
of isolated photon yields and correlations of isolated photons and hadrons in pp and p-Pb collisions. 
The FoCal is composed of different sections-electromagnetic and hadronic calorimeters. 
The electromagnetic calorimeter (FoCal-E) has a hybrid Si+W sandwich design and is crucial for precise  $\gamma/\pi^0$ separation. 
The hadronic sampling calorimeter (FoCal-H) is a spaghetti type calorimeter and is needed for particle identification and jet measurements. 
Ideally, both FoCal-E and FoCal-H should cover the same region of pseudorapidity. 
The distance from the nominal beam interaction point to the FoCal is $\approx$7 m. 
In this way the detector is far from the high particle density flux near the primary interaction vertex and at the same time is close to the beam pipe covering large pseudorapidity \cite{ALICE:2020mso}. 

The construction the FoCal detector requires precise choices of appropriate
technologies which can only be verified in a series of prototype developments
and testing.

%%%%%%%%%%%%%%%%%%%%%%%%%%%%%%%%%%%%%%%%%%
\section{FoCal-H Prototype Design and Readout Electronics} 

\begin{figure}[H]
%\centering
\includegraphics[height=150pt, width=0.32\linewidth]{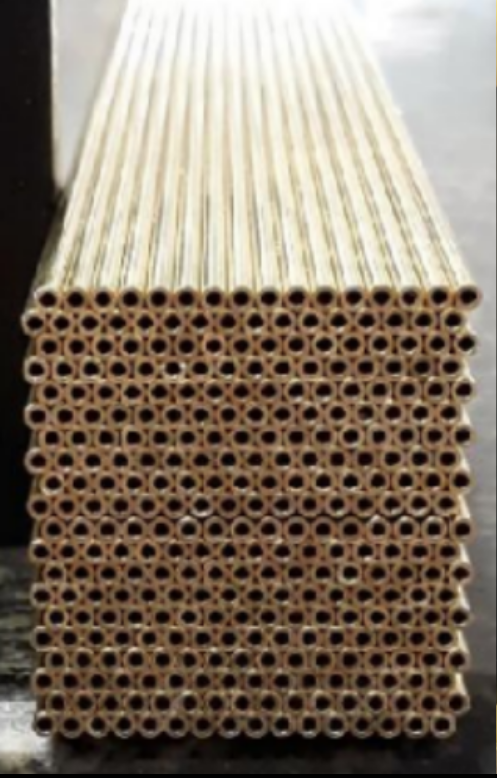}
\includegraphics[height=150pt, width=0.32\linewidth]{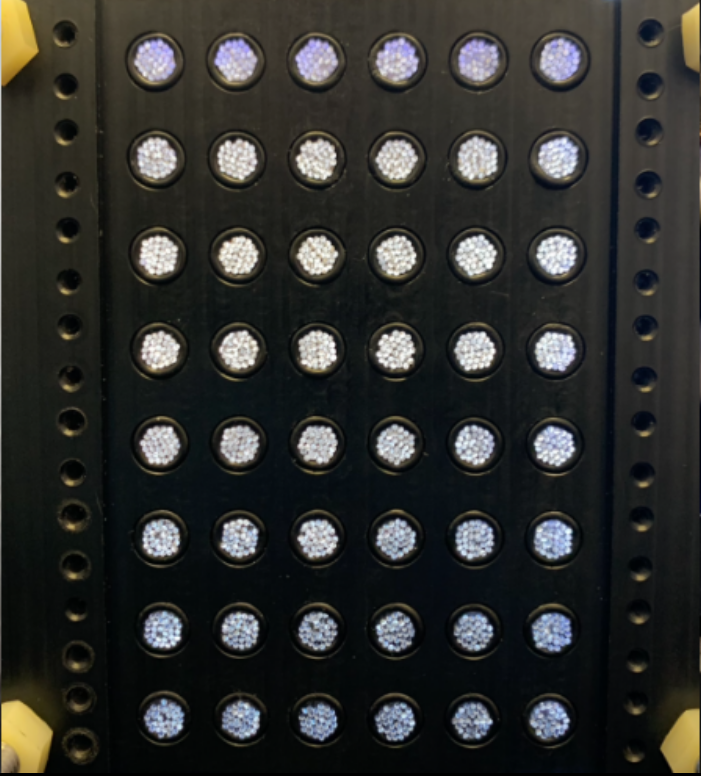}
\includegraphics[height=150pt, width=0.32\linewidth]{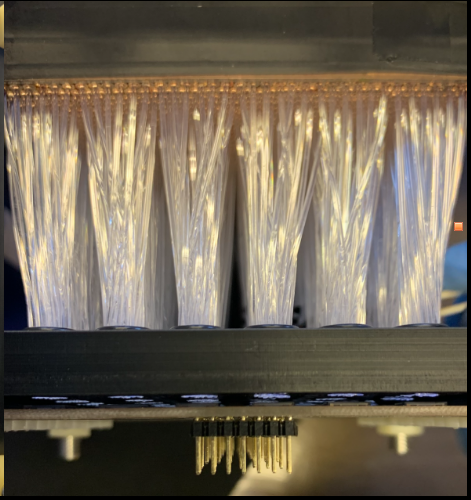}
\caption{ FoCal-H 2021 prototype module with 1440 Cu tubes (\textbf{left}), grouped scintillating fibers at the back of the detector (\textbf{center}) and 16 (out of 48) mounted Onsemi MICROFC-60035-SMT-TR1 SiPMs at the back of the detector (\textbf{right}).} 
\label{fig:detector}
\end{figure}

 The FoCal-H 2021 prototype is a quadrangular prism   calorimeter with dimensions $ 95 \times  95 \times 550$ mm$^3$. The absorber part consists of 
 {$40 \times 36 $  for a total of} 
 1440 Cu capillary tubes, each with \SI{1.2}{\milli\metre} inner and \SI{2.5}{\milli\metre} outer diameter. Inside the tubes a scintillating fiber with \SI{1}{\milli\metre}  diameter is placed. 
% \added{This calorimeter design is known as spaghetti type.}  
 {The downstream end of the fibers are grouped into 48~bundles, }
 each consisting of a 5 $\times$ 6 grid of fibers. 
 %\added{By grouping in this way, enough amount of collected light could be propagated to a corresponding read out system.} 
 {This grouping pattern allowed to obtain equal number of fibers per bundle with transverse dimension matching the photodetector active surface.}
 At the end of each of these groups an Onsemi MICROFC-60035-SMT-TR1 SiPM with a \SI{35}{\micro\metre} cell and active surface of {6 $\times$ 6} mm$^2$ \cite{SiPM} is placed, for a total of 48 channels. An image of the FoCal-H 2021 prototype is given in Figure~\ref{fig:detector}.
 
 The readout of the FoCal-H prototype was based on the CAEN A1702  {\cite{A1702}}
 general purpose board.

The board provides high voltage to each silicon photomultiplier. 
A common positive bias voltage is supplied on the cathode with the possibility to adjust the individual 
channels within a 5 V range by controlling the zero level of the anode. 

 The output signals from the SiPMs are taken from the anode and  are processed by a Weeroc CITIROC ASIC. A 12-bit ADC inside a NXP LPC4370 ARM micro-controller chip \cite{LPC} digitizes the collected charge in each channel \cite{Auger:2016vpo}. The fully-digitized data is transmitted via Ethernet.  
 A modified version of the vendor provided ROOT-based software  {FEBDAQMULT~\cite{A1702}}  was used for the data acquisition. For the present analysis, data taken at equal gain and high voltage settings were selected. 
 
 Two boards were used during the 2021 test. The boards provided signals for two conditional regions. The first one is the central zone of the detector where the beam spot was placed. This region corresponded to a single board (B1) with all 32 active channels and it was the main focus of the 2021 studies. For the second  region, which is the outer zone of the detector, there was the second board (B2) with 16 out of 32 active channels.

%%%%%%%%%%%%%%%%%%%%%%%%%%%%%%%%%%%%%%%%%%
\section{Test Beam Setup }
The constructed prototype was exposed to a  {non-separated}  {secondary} charged beam  {containing electrons, muons and hadrons} with energy 20, 30, 40, 60 and 80 GeV during a two-week test beam at the CERN SPS H6 beam line. In this period various   configurations of the FoCal system were tested. 

\begin{figure}[H]
%\centering
\includegraphics[width=0.8\linewidth]{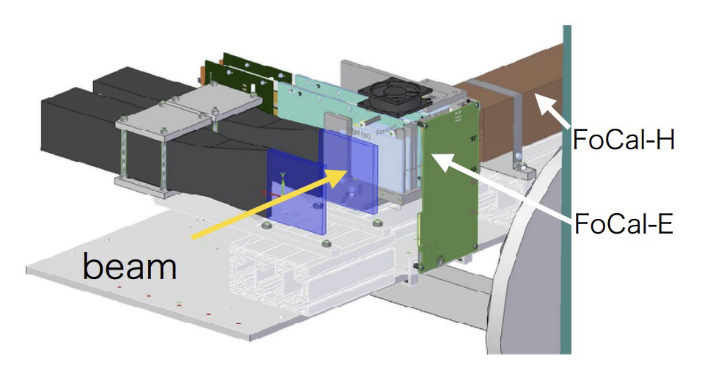}

\caption{A schematic layout of FoCal configuration for data acquisition during TestBeam 2021. } 
\label{fig:set-up}
\end{figure}

The FoCal-H 2021 prototype was mounted on an adjustable basis, allowing us to precisely place the beam spot at the central part of the detector. The distance between the FoCal-H and FoCal-E 2021 prototypes in the test set up was kept at minimum in order to prevent blow-up of the showers from FoCal-E.
A layout of the FoCal-H 2021 prototype placed on the beam line is given in Figure~\ref{fig:set-up}.

\section{Beam Data Analysis}

\subsection{Charge Reconstruction}

During the 2021 test beam, the FoCal-H 2021 prototype collected data with events for beam particles with energies 20, 30, 40, 60 and 80 GeV. For the analysis, a reconstruction environment was developed using ROOT  {\cite{ROOT}} based software. For each channel a pedestal was determined from events in which no signal from the SiPMs was expected. Such events are recorded by providing an  {external}   signal to any of the two trigger inputs{, T0 and T1,} of 
   {the readout board}.  
 {The signal on T0  was with rate 1Hz, while the signal on T1 was with rate 10~Hz, with a constant offset of $\sim$25 ms between them. Since no beam signals were expected in these additional events, they were used as pedestal calculations.} 
  The total   charge was reconstructed as the sum of all pedestal subtracted charges $q_i^k$ in the channels belonging to B1, B2 or both.
 
%\added{
% \begin{equation}
%    Q_1 = \sum_{i=1}^{32} q_i^1 ~~; ~~~ Q_2 = \sum_{j=1}^{16} q_j^2~~; ~~~ Q_{all} = Q_1 + Q_2.
%\end{equation}
%}

Due to problems with boards synchronization only the total charge   using B1 channels was used in the subsequent analysis.

\begin{figure}[H]
%\centering
\includegraphics[width=0.8\linewidth]{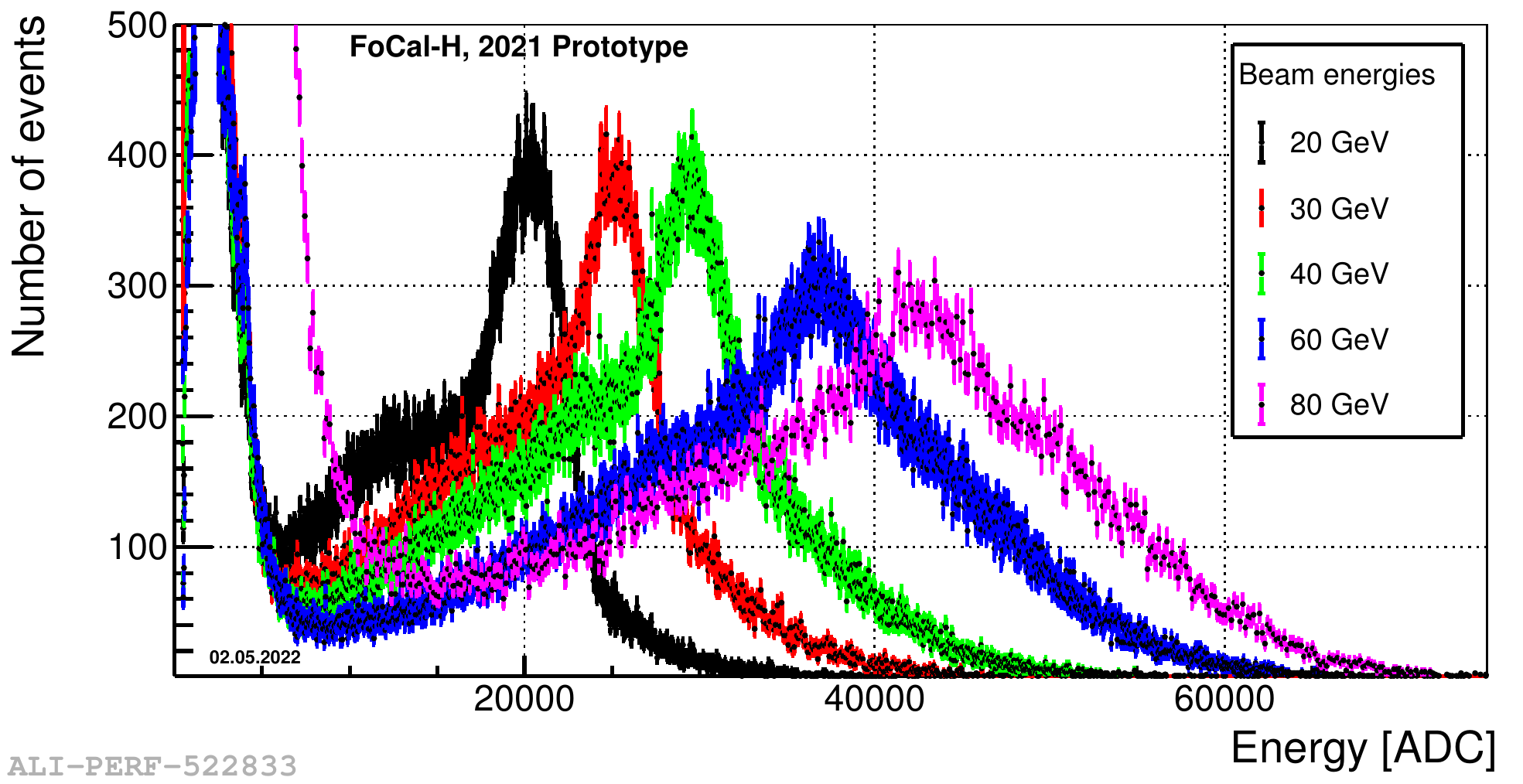}
\caption{Reconstructed %MDPI: Please use commas to separate thousands for numbers with five or more digits (not for four digits) in the picture. e.g., "10000" should be "10,000"
%Radoslav Simeonov: This is how the plot was approved by the ALICE collaboration and this is the only way it can be used in the paper
 charge in the FoCal-H prototype in ADC counts as a function of the beam particles energy for beam energies from 20 GeV to 80 GeV.} 
\label{fig:charge}
\end{figure}

The reconstructed charge in ADC counts as a function of the beam energy is shown in Figure~\ref{fig:charge}.  The MIP peak (on the left) is at the same position for different beam energies, while the second peak is dependant on the beam energy.   

\subsection{Beam Angle Dependence}

When the detector is placed parallel to the beam axis, a particle from the beam can travel through a non-absorber medium in which case it does not start a shower in the~calorimeter.

When this particle travels along a scintillation fiber, the deposited energy and the produced scintillation light lead to a saturation in the 12-bit ADC and results in a peak in the total deposited charge at the position of 4096 ADC counts.

\begin{figure}[H]
%\centering
\includegraphics[width=0.8\linewidth]{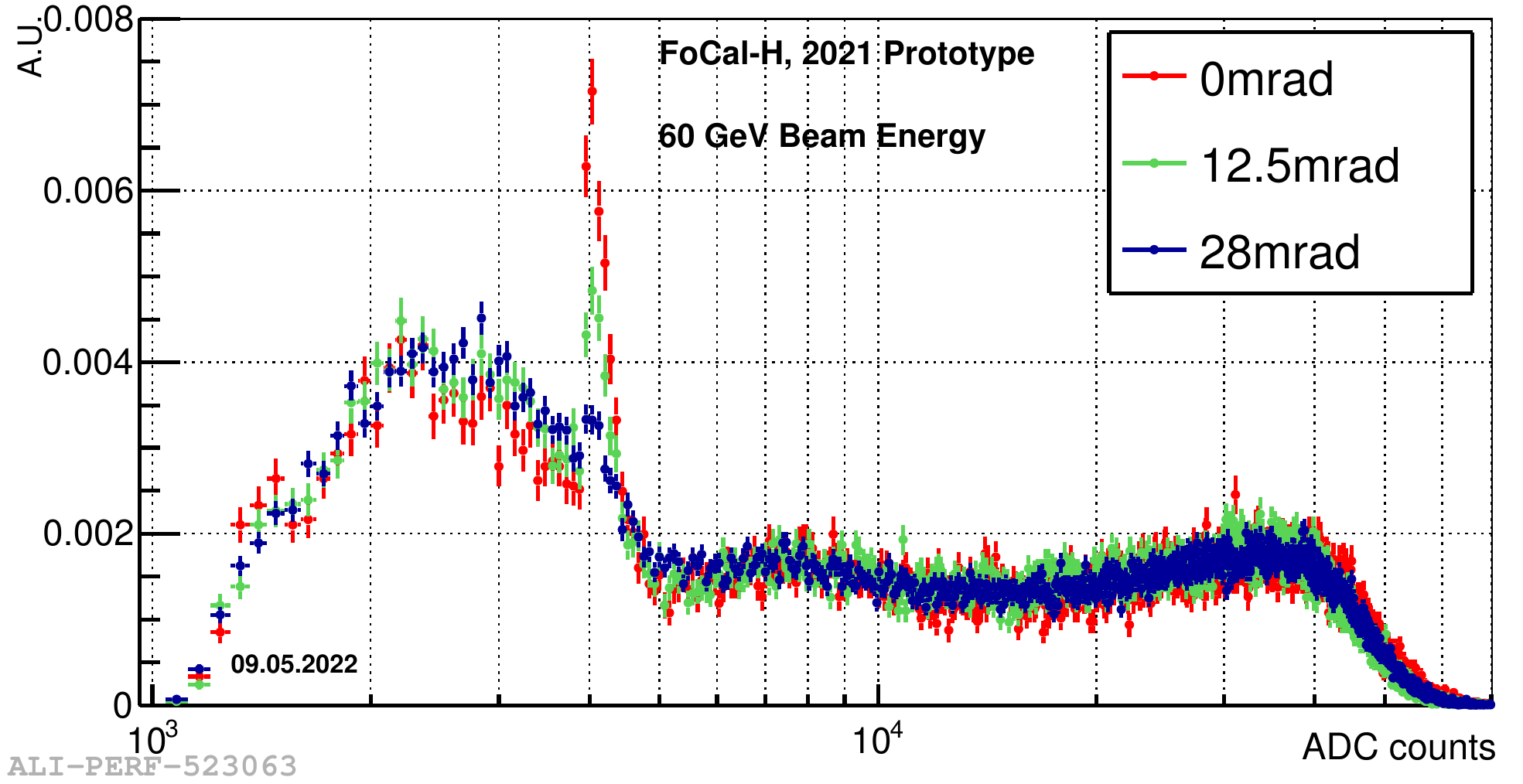}
\caption{The peak corresponding to a particle traveling along the fiber decreases with the increase of the angle  {between the detector and the beam axis} and almost vanishes at 28~mrad.  {For the subsequent results data taken with a spacer providing 28~mrad inclination was used.}}

\label{fig:angle}
\end{figure}

To decrease this effect a small spacer was added at the end of the FoCal-H 2021 prototype. By this, a non-zero angle is achieved between the axis of the beam and the longitudinal axis of the detector. Two different spacers were used to study the effect of lowering the saturation in the ADC. Their dimensions of \SI{5.5}{\milli\metre} and \SI{11}{\milli\metre} provide \mbox{12.5 mrad} and 28 mrad inclination of the detector. The total charge distribution dependence on the angle of incidence from the FoCal-H 2021 prototype data is given in Figure~\ref{fig:angle}.  {The single channel saturation peak decreases significantly for angle of 28 mrad and for the subsequent results data taken with the \SI{11}{\milli\metre} spacer was used.}

\section{Monte-Carlo Simulations}

A full Monte-Carlo (MC) simulation of the passage of particles through the detector volume was performed using the GEANT4 toolkit \cite{Geant}. 
The design and the geometry of the FoCal-H 2021 
prototype were implemented and precise studies of the 
    {prototype response}
were performed. 
The simulation of the read out electronics was  {not} implemented. 
 {A single parameter was introduced to account 
for the conversion from deposited energy to collected charge }
as part of the analysis of the simulated data. 
The physics list used for the simulations was FFTP\_BERT, 
but some of the results were also checked with QGSP\_BERT. 
The energy deposit in each plastic scintillator fiber was taken as an output from the simulation, 
while light propagation, 
SiPM response and digitization    {by converting 
MeV to pC was applied} at the analysis level.   

One of the main goals for the MC was to describe the total charge distribution containment of the data. Simulations with the same number of events for pions ($\pi$), muons ($\mu$), protons (p) and electrons (e) were performed for simulation beam energies of 40 GeV and 60 GeV. The following total charge distributions were summed with weights, where the total amount of weight of each particle distribution was computed by using the Atherton parameterization \cite{Atherton:1980vj}.  {This parameterization gives the fluxes of the different type hadrons in the secondary beam produced in a p-Be collisions.}

\begin{figure}[H]
%\centering
\includegraphics[width=0.8\linewidth]{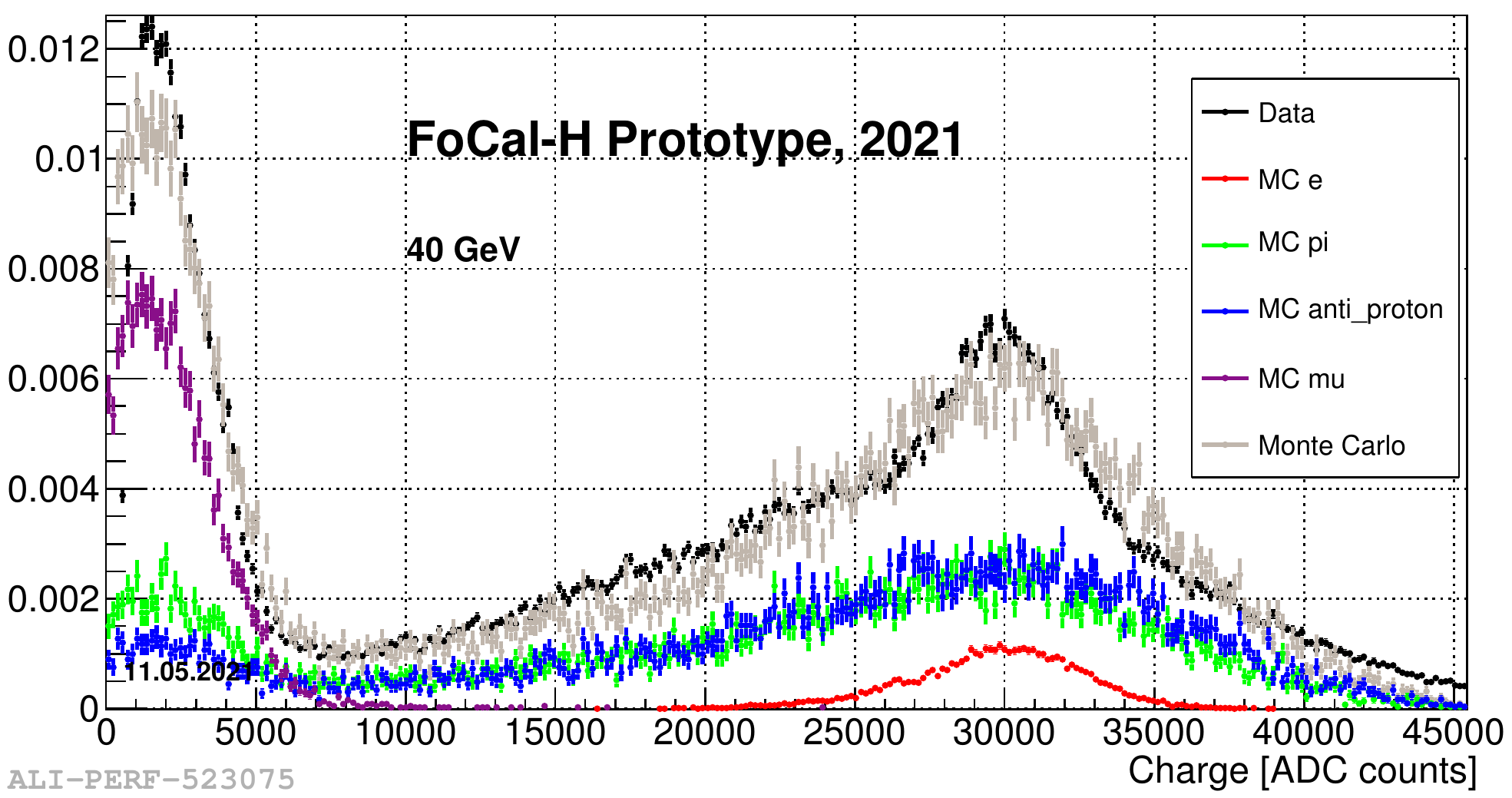}
\includegraphics[width=0.8\linewidth]{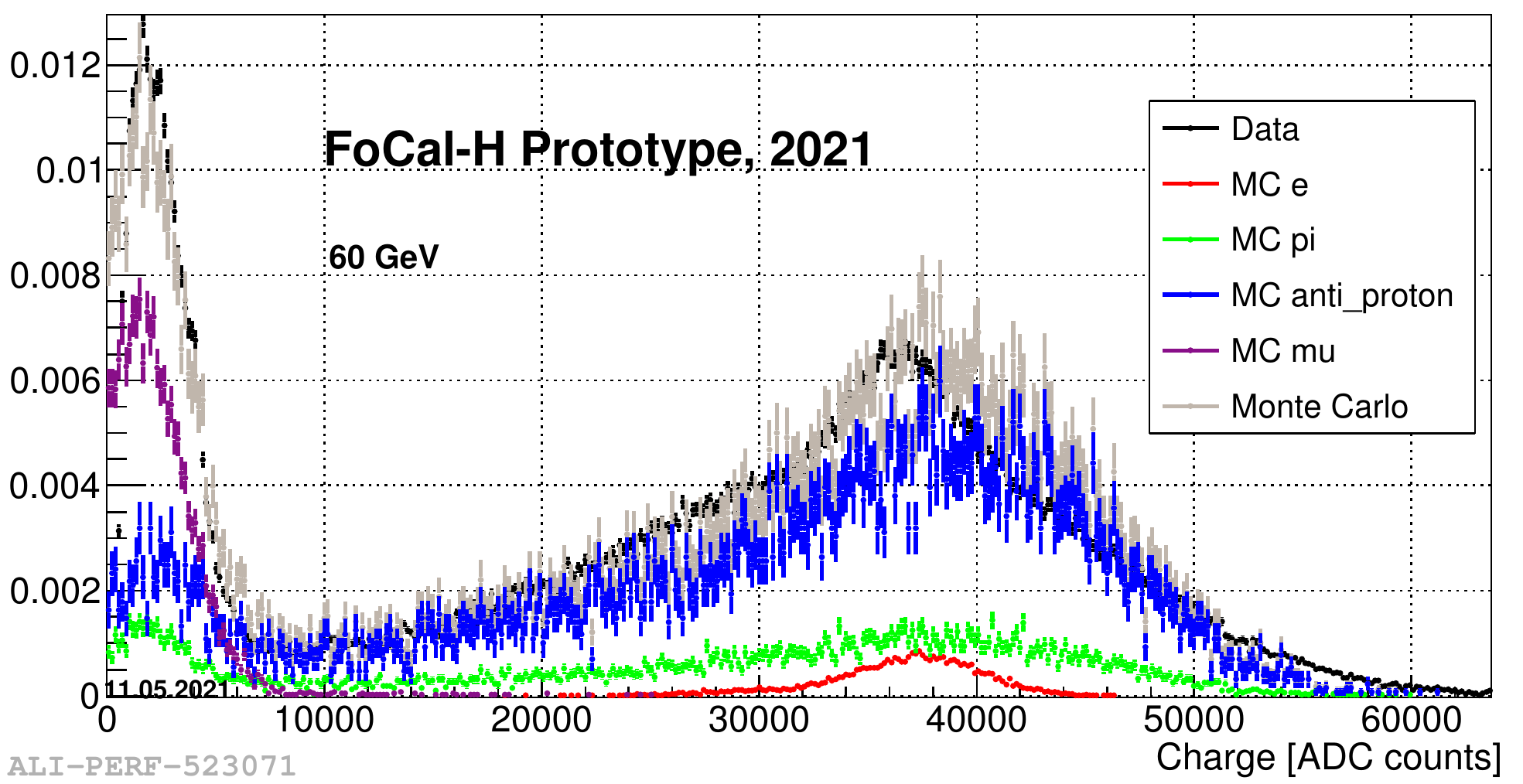}

\caption{Total %MDPI: Please use commas to separate thousands for numbers with five or more digits (not for four digits) in the picture. e.g., "10000" should be "10,000"
%Radoslav Simeonov: This is how the plot was approved by the ALICE collaboration and this is the only way it can be used in the paper
 charge distribution in data compared with a weighted sum of MC simulated events with 40 GeV (\textbf{top}) and 60 GeV (\textbf{bottom}) e, $\pi$, $\mu$, p.} 

\label{fig:40_GeV}

\end{figure}

The weighted sum of these distributions for each of these two energies was compared with the total charge distribution from the data. The data total charge distributions described by a weighted sum of Monte-Carlo simulated beam particles for 40 GeV and \mbox{60 GeV} are given in Figure~\ref{fig:40_GeV}.

A comparison of the dependence of the reconstructed charge from different beam particle energies between the data and the MC was performed for energies from 20 to \mbox{80 GeV}. {To estimate the reconstructed charge from the data, we assumed that the right peak in total energy distribution is due to electrons and we took the value of the charge after we fit it.}  For the  simulation dedicated electron simulations were executed for the same energies. Again, the electron peak was determined from the total energy distribution in the simulated detector environment. 

\begin{figure}[H]
%\centering
\includegraphics[width=0.75\linewidth]{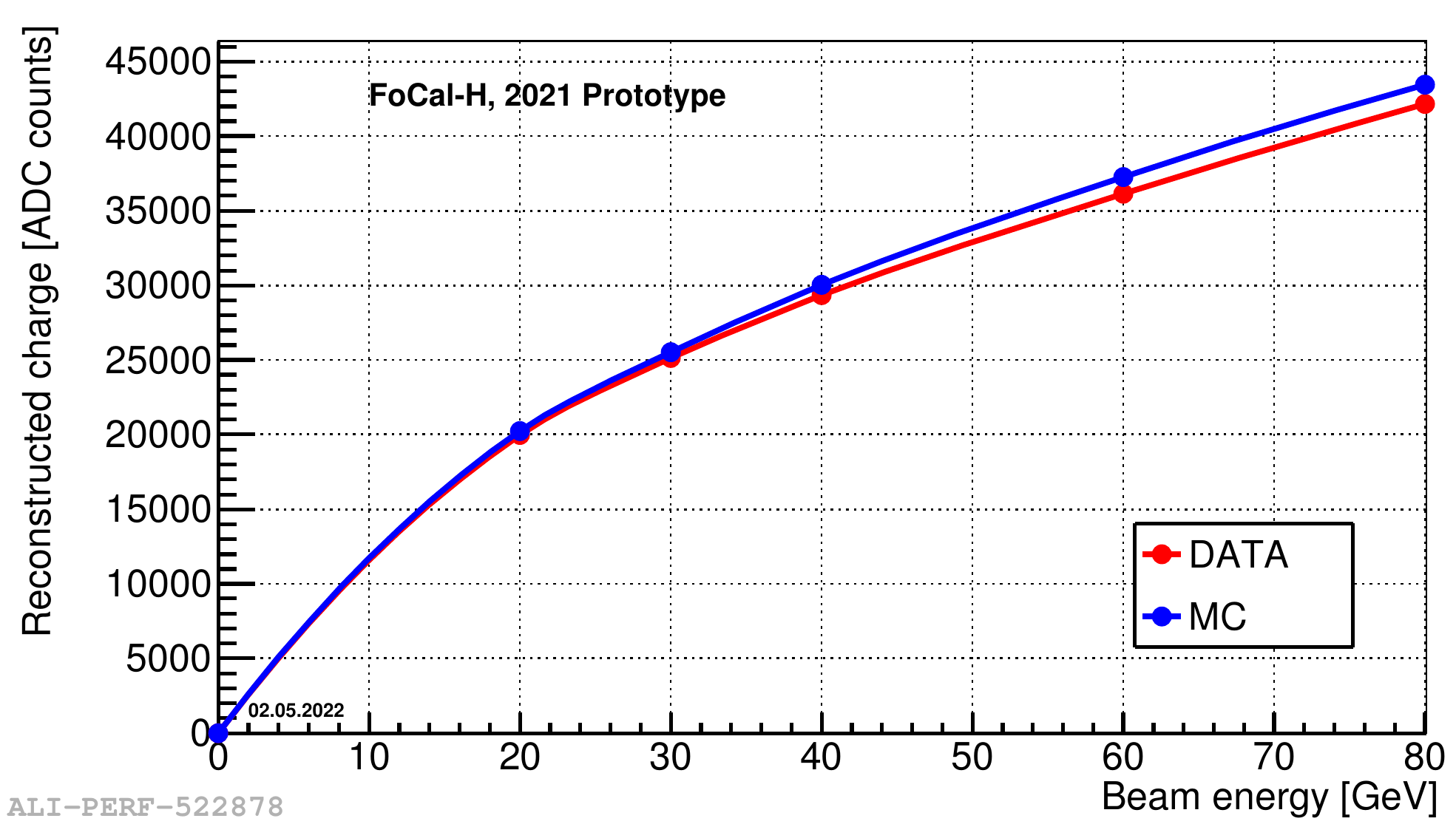}
\caption{Dependence %MDPI: Please use commas to separate thousands for numbers with five or more digits (not for four digits) in the picture. e.g., "10000" should be "10,000"
%Radoslav Simeonov: This is how the plot was approved by the ALICE collaboration and this is the only way it can be used in the paper
 of the reconstructed electron charge (peak value) in ADC counts on beam particle energies from 20 to 80 GeV.} 
\label{fig:charge_reco}
\end{figure}

Dependence of the reconstructed electron charge (peak value) in ADC counts on the beam energy can be seen on Figure~\ref{fig:charge_reco}.

\vspace{-6pt}

%%%%%%%%%%%%%%%%%%%%%%%%%%%%%%%%%%%%%%%%%%

The results of the simulation were found to be consistent with the data acquired during the test beam. The successful Monte Carlo implementation will also be used in the further development of FoCal-H prototypes including the final detector. 

\section{Conclusions}

The FoCal detector consisting out of an electromagnetic and hadronic calorimeter is a planned upgrade of ALICE for Run 4. The first prototype of the hadronic subsystem was successfully assembled and tested on the H6 beamline at the CERN SPS.  Beam energy scanning was performed in the interval 20--80 GeV and the event reconstruction followed qualitatively the expected trend. A Monte-Carlo model was developed and the results showed consistency with the acquired data.  A larger prototype is currently under construction and improved performance is expected by the end of 2022.

%%%%%%%%%%%%%%%%%%%%%%%%%%%%%%%%%%%%%%%%%%
\vspace{6pt} 

%%%%%%%%%%%%%%%%%%%%%%%%%%%%%%%%%%%%%%%%%%
%\authorcontributions{The presented material is a result of the joint work of the ALICE collaboration. All authors have read and agreed to the published version of the manuscript. 
%Please add  Author Contributions: For research articles with several authors, a short paragraph specifying their individual contributions must be provided. If there is “Author Contributions” section, the group/institution or its author’s contribution should be mentioned. 

\funding{The work is partly supported by Carlsberg Foundation CF21-0606, Hadronic Calorimeter for Forward Physics, Bulgarian National Roadmap for Research Infrastructures-CERN D01-374/18.12.2020 and Bulgarian National Program Young Scientists and Postdoctoral Fellows \mbox{(MUPD-2).}}
%\institutionalreview{\hl{    }}%In this section, please add the Institutional Review Board Statement and approval number for studies involving humans or animals. Please note that the Editorial Office might ask you for further information. Please add “The study was conducted according to the guidelines of the Declaration of Helsinki, and approved by the Institutional Review Board (or Ethics Committee) of NAME OF INSTITUTE (protocol code XXX and date of approval).” OR “Ethical review and approval were waived for this study, due to REASON (please provide a detailed justification).” OR “Not applicable” for studies not involving humans or animals. You might also choose to ex-clude this statement if the study did not involve humans or animals.
%\informedconsent{\hl{      }}%Any research article describing a study involving humans should contain this statement. Please add “Informed consent was obtained from all subjects involved in the study.” OR “Patient con-sent was waived due to REASON (please provide a detailed justification).” OR “Not applicable” for studies not involving humans. You might also choose to exclude this statement if the study did not involve humans.
%Written informed consent for publication must be obtained from participating patients who can be identified (including by the patients themselves). Please state “Written informed consent has been obtained from the patient(s) to publish this paper” if applicable.

\dataavailability{ Not applicable.} %In this section, please provide details regarding where data supporting reported results can be found, including links to publicly archived datasets analyzed or generated during the study. Please refer to suggested Data Availability Statements in section “MDPI Research Data Policies” at \href{https://www.mdpi.com/ethics}{https://www.mdpi.com/ethics}. You might choose to exclude this statement if the study did not report any data.

%\acknowledgments{}

%\acknowledgments{The work is partly supported by Carlsberg Foundation CF21-0606, Hadronic Calorimeter for Forward Physics, Bulgarian National Roadmap for Research Infrastructures-CERN D01-374/18.12.2020 and Bulgarian National Program Young Scientists and Postdoctoral Fellows (MUPD-2).}
\conflictsofinterest{The authors declare no conflict of interest. %Please disclose any conflicts of interest, or add “The authors declare no conflicts of interest.
%Radoslav Simeonov: Done
}

%%%%%%%%%%%%%%%%%%%%%%%%%%%%%%%%%%%%%%%%%%
\begin{adjustwidth}{-\extralength}{0cm}

\reftitle{References}

%=====================================

%%%%%%%%%%%%%%%%%%%%%%%%%%%%%%%%%%%%%%%%%%
%% for journal Sci
%\reviewreports{\\
%Reviewer 1 comments and authors’ response\\
%Reviewer 2 comments and authors’ response\\
%Reviewer 3 comments and authors’ response
%}
%%%%%%%%%%%%%%%%%%%%%%%%%%%%%%%%%%%%%%%%%%
\end{adjustwidth}
\end{document}